\documentclass[3p,12pt, authoryear,english]{elsarticle}

\usepackage[english]{babel}
\usepackage[utf8]{inputenc}
\usepackage{graphicx}
\usepackage{url}
\usepackage{datetime}
\usepackage{lineno}
\usepackage{amsmath}
\usepackage{booktabs}
\usepackage{algorithm} 
\usepackage{algpseudocode}
\usepackage{breakcites}
\usepackage{caption}
\usepackage{subcaption}
\usepackage{setspace}

\begin{document}

\begin{frontmatter}

\title{MSTL: A Seasonal-Trend Decomposition Algorithm for Time Series with Multiple Seasonal Patterns}

\author[unimelb]{Kasun~Bandara \corref{cor1}}
\author[monash_bus]{Rob J Hyndman}
\author[monash]{Christoph~Bergmeir}

\address[unimelb]{School of Computing and Information Systems, Melbourne Centre for Data Science, University of Melbourne}
\address[monash_bus]{Department of Econometrics and Business Statistics, Monash University}
\address[monash]{Department of Data Science and AI, Monash University}

\cortext[cor1]{Corresponding Author Name: Kasun Bandara, Affiliation: School of Computing and Information Systems, Melbourne Centre for Data Science, University of Melbourne, Melbourne, Australia, Postal Address: School of Computing and Information Systems, The University of Melbourne, Victoria 3052, Australia, E-mail address: Kasun.Bandara@unimelb.edu.au}

\doublespacing

\begin{abstract}
The decomposition of time series into components is an important task that helps to understand time series and can enable better forecasting. Nowadays, with high sampling rates leading to high-frequency data (such as daily, hourly, or minutely data), many real-world datasets contain time series data that can exhibit multiple seasonal patterns. Although several methods have been proposed to decompose time series better under these circumstances, they are often computationally inefficient or inaccurate. In this study, we propose Multiple Seasonal-Trend decomposition using Loess (MSTL), an extension to the traditional Seasonal-Trend decomposition using Loess (STL) procedure, allowing the decomposition of time series with multiple seasonal patterns. In our evaluation on synthetic and a perturbed real-world time series dataset, compared to other decomposition benchmarks, MSTL demonstrates competitive results with lower computational cost. The implementation of MSTL is available in the R package \textit{forecast}.
\end{abstract}

\begin{keyword}
Time Series Decomposition, Multiple Seasonality, MSTL, TBATS, STR
\end{keyword}

\end{frontmatter}

\doublespacing
\section{Introduction}
In time series analysis and forecasting, it is often useful to identify the underlying patterns of time series data, to better understand the contributing phenomena and to enable better forecasting.
Time series decomposition techniques have been introduced to decompose a time series into multiple components including trend, seasonality, and remainder. Such methods have important time series applications in seasonal adjustment procedures~\citep{Maravall2006-ty, Thornton2013-ir}, forecasting~\citep{Koopman2006-ot,Hewamalage2020-ko, Bandara2020-zt} and anomaly detection~\citep{Wen2019-fk, Wen2020-rv}.

\begin{figure}[htb]
 \centering
\includegraphics[width=0.8\textwidth]{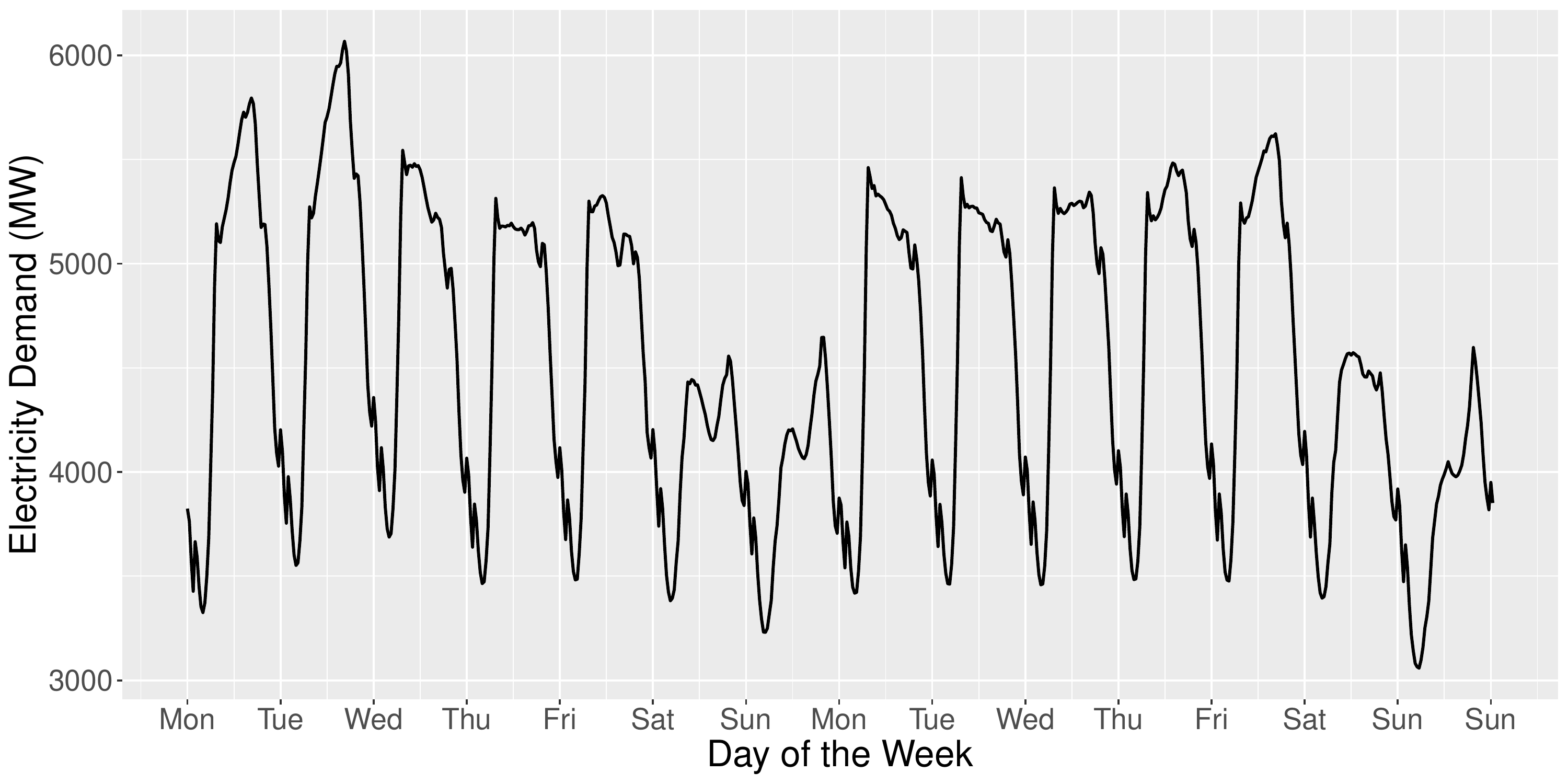}
 \caption{Half-hourly energy demand in the state of Victoria, Australia over a two weeks period of time, extracted from the \textit{vic\_elec} dataset \protect \citep{OHara-Wild2020-uq} displaying daily and weekly seasonal patterns.}
\label{fig:multiseasonalplot}
\end{figure}

Nowadays, with rapid growth of availability of sensors and data storage capabilities, there is a significant increase of time series with higher sampling rates (sub-hourly, hourly, daily). Compared with traditional time series data, the higher-frequency time series data may exhibit more complex properties, such as multiple seasonal cycles, non-integer seasonality, etc. These time series are commonly found in the utility demand industry (electricity and water usage), mainly due to the intricate usage patterns of humans. For example, Figure ~\ref{fig:multiseasonalplot} illustrates the aggregated half-hourly energy demand time series in the state of Victoria, Australia that has two seasonal periods, a daily seasonality (period $= 48$) and a weekly seasonality (period $= 336$). It can be seen that the average energy consumption in the weekdays is relatively higher to those in the weekends. Furthermore, a longer version of this time series may even show a yearly seasonality (period $= 17532$), with values systematically changing across the seasons such as summer and winter. In this scenario, daily and weekly consumption patterns are useful to estimate the short-term energy requirements; the yearly seasonal usage patterns are beneficial in long-term energy planning. Therefore, the accurate decomposition of time series with multiple seasonal cycles is useful to provide better grounds for decision-making in various circumstances.

There is a large body of literature available on time series decomposition techniques. Widely used traditional time series decomposition methods are Seasonal-Trend decomposition using Loess \citep[STL,][]{Cleveland1990-rc}, X-13-ARIMA-SEATS \citep{Bell1984-vo}, and X-12-ARIMA \citep{Findley1998-cm}. Although these methods have been heavily used in many real-world applications due to their robustness and efficiency, these techniques can handle only time series with a single seasonality. More recently, methods to decompose time series with multiple seasonal patterns have been introduced \citep{str, Wen2020-rv}. For example,~\cite{str} introduced Seasonal-Trend decomposition by Regression (STR), a regression based, additive decomposition technique, which is also capable of modelling the influence of external factors towards the seasonal patterns in a time series. ~\cite{Wen2020-rv} recently developed Fast-RobustSTL, a decomposition technique that accounts for multiple seasonal patterns and noise in time series. Several studies also support the use of forecasting models to extract seasonal patterns from time series~\citep{Bandara2020-zt, Bandara2020-jn, Bandara2021-pattern}. The idea is to fit a forecasting model to the time series that is capable of handling time series with multiple seasonal patterns, such as TBATS (Trigonometric Exponential Smoothing State Space model with Box-Cox transformation, ARMA errors, Trend and Seasonal Components)~\citep{De_Livera2011-sg}, and Prophet~\citep{prophetpackage}; thereafter, to extract the fitted time series components, i.e., trend, multiple seasonal components, from the fitted forecast model. Nonetheless, the objective function of these forecast models is to minimise the prediction error and therefore these methods have to limit themselves to only past data for decomposition, whereas dedicated decomposition methods can use both past and future information. Therefore, the use of prediction based approaches for time series decomposition can be inaccurate and may give not the best possible decomposition of a time series.

In this paper, we introduce Multiple STL Decomposition (MSTL), a fully automated, additive time series decomposition algorithm to handle time series with multiple seasonal cycles. The proposed MSTL algorithm is an extended version of the STL decomposition algorithm, where the STL procedure is applied iteratively to estimate the multiple seasonal components in a time series. This allows MSTL to control the smoothness of the change of seasonal components for each seasonal cycle extracted from the time series, and seamlessly separate their seasonal variations (e.g., deterministic and stochastic seasonalities). For non-seasonal time series, MSTL determines only the trend and remainder components of the time series.

Specifically, the MSTL algorithm initially determines the number of distinct seasonal patterns available in the time series. Often times, the multiple seasonal patterns are structurally unnested and interlacing together. As a result, during decomposition, the seasonal components relevant to a lower seasonal cycle can be excessively absorbed by a higher seasonal cycle. To minimise such seasonal confounding, as the second step, MSTL arranges the identified seasonal cycles in an ascending order. Then, if the time series is seasonal, MSTL applies the STL algorithm iteratively to each of the identified seasonal frequencies. Next, the trend component of the time series is computed using the last iteration of STL. On the other hand, if the time series is non-seasonal, MSTL uses the Friedman's Super Smoother function, \textit{supsmu}, available in R \citep{Rcore}, to directly estimate the trend of the time series. Finally, to calculate the remainder part of seasonal time series, the trend component is subtracted from the seasonally adjusted time series. Whereas, for non-seasonal time series, the trend component is subtracted from the original time series to derive the remainder.

\begin{algorithm}
\begin{algorithmic}[1]
\Procedure{MSTL}{X, iterate, $\lambda$, s.win[], \dots}
\If{X is multi-seasonal}
\State seas.ids[] $\gets$ attributes(X)
\State k $\gets$ len(X)
\State seas.ids $\gets$ sea.ids[seas.ids $<$ k/ 2]
\State seas.ids $\gets$ sort(seas.ids, dec = F)
\ElsIf{X is single-seasonal}
\State seas.ids[] $\gets$ frequency(X)
\State iterate $\gets$ 1
\EndIf
\State X $\gets$ \textbf{na.interp}(X, $\lambda$) 
\State X $\gets$ \textbf{BoxCox}(X, $\lambda$)
\If{seas.ids[1] $>$ 1}
\State seasonality $\gets$ list(rep(0, len(seas.ids)))
\State deseas $\gets$ X
\For{j in 1 to iterate}
\For{i in 1 to len(seas.ids)}
\State deseas $\gets$ deseas + seasonality[[i]]
\State fit $\gets$ \textbf{STL}(ts(deseas, frequency = seas.ids[i]), s.window = s.win[i], \dots)
\State seasonality[[i]] $\gets$ msts(seasonal(fit))
\State deseas $\gets$ deseas - seasonality[[i]]
\EndFor
\EndFor
\State trend $\gets$ msts(trendcycle(fit))
\Else 
\State seas.ids$\gets$ NULL
\State deseas $\gets$ X
\State trend $\gets$ ts(\textbf{SUPSMU}(X))
\EndIf 
\State remainder $\gets$ deseas - trend
\State \textbf{return} [trend, remainder, seasonality] 
\EndProcedure
\caption{MSTL decomposition.}
\label{alg:mstl}
\end{algorithmic}
\end{algorithm}

MSTL is a robust, accurate seasonal-trend decomposition algorithm that is designed to capture multiple seasonal patterns in a time series. Most importantly, compared with other decomposition alternatives, MSTL is an extremely fast, computationally efficient algorithm, which is scalable to increasing volumes of time series data. In R, the proposed MSTL algorithm is implemented in the \verb|mstl| function from the \textit{forecast} package \citep{forecastpackagepaper,forecastpackage}.

\section{Model Overview}
Similar to the STL algorithm, MSTL gives an additive decomposition of the time series. Given $X_t$ is the observation at time $t$, the additive decomposition can be defined as follows:
\begin{equation}
X_t = \hat{S}_t + \hat{T}_t + \hat{R}_t
\label{additive}
\end{equation}
Here, $\hat{S}_t $, $\hat{T}_t$, $\hat{R}_t$ denote the seasonal, trend, and remainder components of the observation, respectively. MSTL extends Equation~\ref{additive} to include multiple seasonal patterns in a time series as follows:
\begin{equation}
X_t = \hat{S}^1_t + \hat{S}^2_t + \dots + \hat{S}^n_t + \hat{T}_t + \hat{R}_t
\label{multiadditive}
\end{equation}
Here, $n$ represents the number of seasonal cycles present in $X_t$. \\

To summarize, a scheme of the MSTL procedure is given in Algorithm~\ref{alg:mstl}. At first, the frequencies of the seasonal patterns in the time series are identified and sorted in an ascending order. Here, the frequencies which are smaller than half of the length of the series are ignored, as those frequencies cannot exhibit any seasonal patterns. After identifying the seasonality, the missing values of the time series are imputed using the \verb|na.interp| function available from the \textit{forecast} package~\citep{forecastpackage}. Next, if $\lambda \in [0,1]$ is given, a Box-Cox transformation is applied to the time series accordingly, using the \verb|BoxCox| function available from the \textit{forecast} package~\citep{forecastpackage}. Thereafter, to every selected seasonal cycle the STL decomposition is fitted. Here, the inner-loop repeats the STL procedure to extract the seasonal components from the time series. The STL model is implemented using the \verb|stl| function provided by the \verb|stats| package in R~\citep{Rcore}. In STL, the rate of seasonal variation is controlled by the \verb|s.window| parameter. Here, a smaller value of \verb|s.window| is set if the seasonal pattern evolves quickly, whereas a higher value is used if the seasonal pattern is constant over time. For example, adjusting the s.window parameter to ``periodic'' limits the change in the seasonal components to zero, which can extract the deterministic seasonality from a time series. In MSTL, a vector of \verb|s.window| can be provided to control the variation of the seasonal components of each seasonal cycle. The outer-loop iterates the STL procedure multiple times to refine the extracted seasonal components. After executing the outer-loop, MSTL calculates the trend component of the time series using the final iteration of STL. In situations where a time series fails the seasonality test, MSTL applies the \texttt{supsmu} function to ascertain the trend of the time series. Finally, the remainder component is retrieved by deducting the trend component from the seasonally-adjusted time series. Also, since MSTL extends the STL algorithm, the other parameters of STL (e.g., \verb|t.window|, \verb|l.window|) are inherited by MSTL, and can be used for decomposition.

\section{Experimental Setup}
We evaluate the proposed MSTL decomposition algorithm on both simulated and a perturbed real-world time series dataset, where we know the true composition, i.e., trend, seasonality, and remainder, of time series. 

\subsection{Benchmarks}
We compare MSTL against a collection of current state-of-the-art techniques in decomposing time series with multiple seasonal cycles. This includes STR~\citep{str} as a pure decomposition technique, TBATS~\citep{De_Livera2011-sg} and Prophet~\citep{prophetpackage} as forecasting techniques, where we use the decomposition of the time series.

\begin{itemize}
\item \textit{STR}: A regression based, additive decomposition technique. In R, the STR algorithm is available through the \verb|STR| function from the stR package~\citep{strpaper2021}.
\item \textit{TBATS}: A state-of-the-art technique to forecast time series with multiple seasonal cycles. TBATS uses trigonometric expression terms to model complex seasonal terms in a time series. In our experiments, we use the R implementation of the TBATS algorithm, \verb|tbats|, from the forecast package~\citep{forecastpackage}.
\item \textit{PROPHET}: An automated forecasting framework, developed by Facebook, that can handle multiple seasonal patterns. Similar to STR and MSTL, Prophet is an additive decomposition technique. In our evaluation, we apply the Prophet algorithm available through the \verb|Prophet| package in R~\citep{prophetpackage}.
\end{itemize}

\subsection{Evaluation metric}

The accuracy of the decomposition methods are evaluated using the root mean square error (RMSE). The RMSE is defined as follows:
\begin{equation}
 RMSE = \sqrt{\frac{1}{n}\sum_{i=1}^{n}(X_{t} - \hat{X_{t}})^{2}}
\label{RMSE}
\end{equation}
Here, $X_{t}$ represents the actual decomposition value at time $t$, and $\hat{X_{t}}$ is the estimated decomposition value. Also, $n$ is the number of observations in the time series.

\subsection{Simulated Data}
\label{sec:simulation}

We simulate daily and hourly data using four time series components. The additive decomposition of the daily time series ($X^{D}_t$) and the hourly time series ($X^{H}_t$) can be formulated as follows:
\begin{align}
X^{D}_t & = T_t + \alpha S^{W}_t + \beta S^{Y}_t + \gamma {R}_t , \quad t=1,\dots,n, \label{daily_additive} \\
X^{H}_t & = T_t + \alpha S^{D}_t + \beta S^{W}_t + \gamma {R}_t , \quad t=1,\dots,m, \label{hourly_additive}
\end{align}

In Equation~\eqref{daily_additive}, \(T_{t}\) is the trend, \(S^{W}_{t}\) is the weekly seasonal component, \(S^{Y}_{t}\) is the yearly seasonal component, and \(R_{t}\) is the remainder of $X^{D}_t$. Whereas in Equation~\eqref{hourly_additive}, \(T_{t}\), \(S^{D}_{t}\), \(S^{W}_{t}\), and \(R_{t}\) corresponds to the trend, daily seasonal component, weekly seasonal component, and remainder component of $X^{H}_t$ respectively. Here, \(\alpha\), \(\beta\), and \(\gamma\) are parameters which control the contribution of the components to $X^{D}_t$ and $X^{H}_t$. Also, $n$ denotes the length of the daily time series and $m$ is the length of the hourly time series.

We simulate the time series data using two data generating processes (DGPs), the ``Deterministic DGP'' and the ``Stochastic DGP"". Here, the Deterministic DGP generates time series that has deterministic components that are invariant to time, whereas the Stochastic DGP gives time series with time-varying components. 

In our experiments, the deterministic trend components \(T_t^{d}\) are generated using a quadratic trend function with random coefficients, \(T_t^{d} = N_1(t + n/2(N_2-1))^2\) where \(N_1\) and \(N_2\) are independent N(0,1) random variables. The deterministic seasonal components \(S_t^{d}\) are composed of five pairs of Fourier terms with random N(0,1) coefficients. Both \(T_t^{d}\) and \(S_t^{d}\) are normalised to give mean zero and unit variance.

To generate stochastic trend components \(T_t^{s}\), we use an ARIMA(0,2,0) model with standard normal errors. With respect to the stochastic seasonal components \(S_t^{s}\), we introduce an additional error term N(0,$\sigma^2$) to \(S_t^{d}\) to change the coefficients for the Fourier terms from one seasonal cycle to another. Here, the stochastic strength of the seasonality is controlled by the $\sigma^2$ parameter value. In other words, the $\sigma^2 = 0$ scenario of \(S_t^{s}\) represents the \(S_t^{d}\).

\begin{figure}
     \centering
     \begin{subfigure}[b]{0.70\textwidth}
         \centering
         \includegraphics[width=\textwidth]{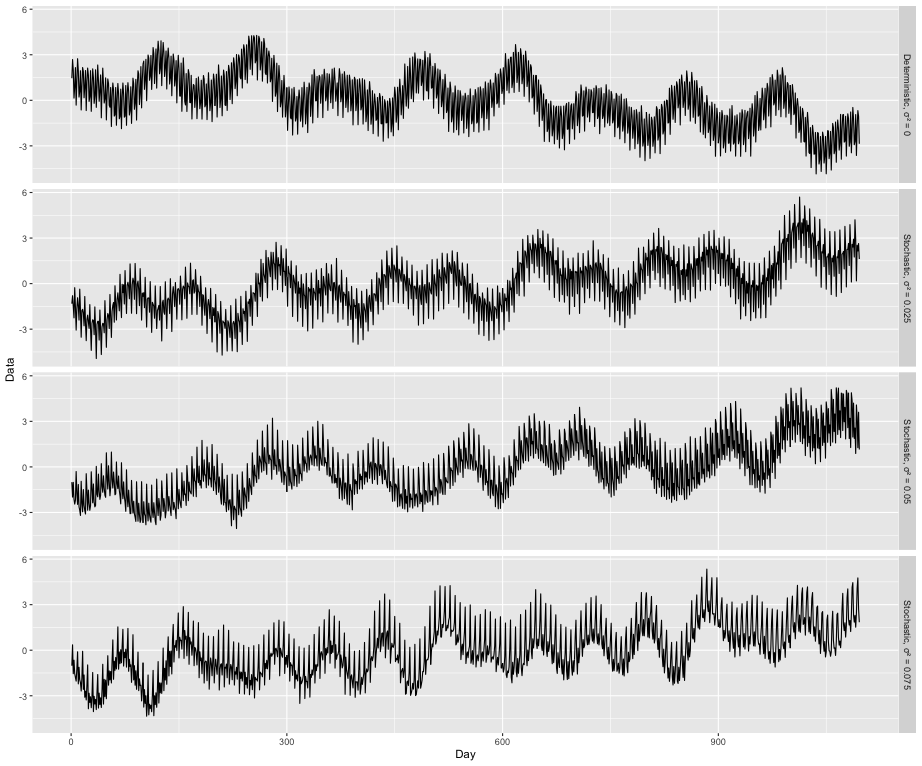}
         \caption{The simulated daily time series with different values of $\sigma^2$  that control the seasonal stochastic components.}
         \label{fig:daily_simulated}
     \end{subfigure}
     \vfill
     \begin{subfigure}[b]{0.70\textwidth}
         \centering
         \includegraphics[width=\textwidth]{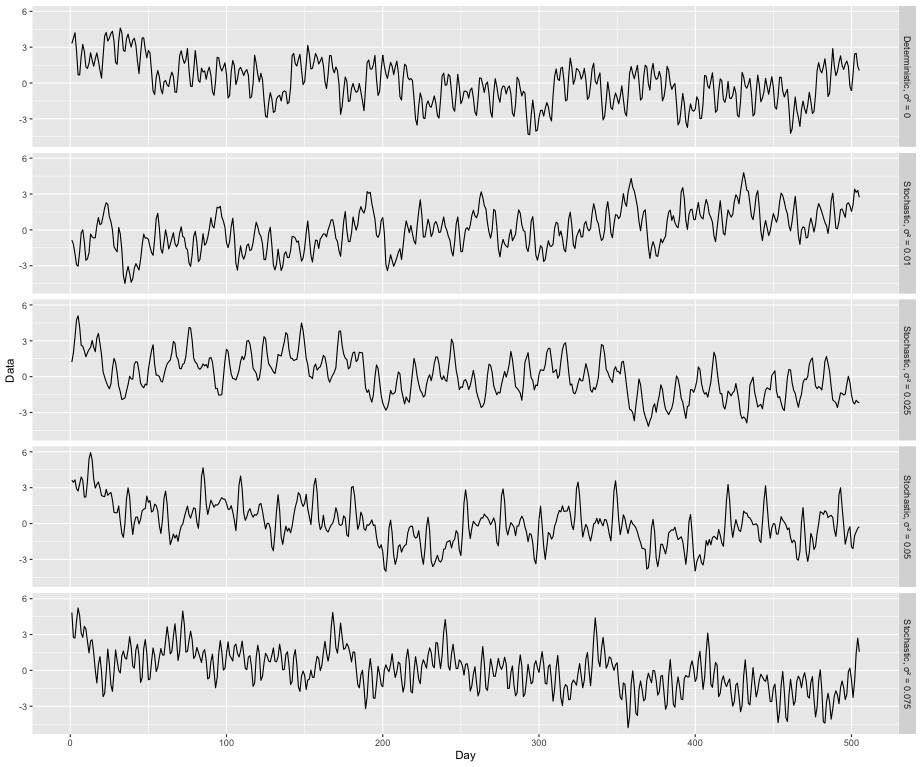}
         \caption{The simulated hourly time series with different values of $\sigma^2$  that control the seasonal stochastic components.}
         \label{fig:hourly_simulated}
     \end{subfigure}
        \caption{Examples of daily and hourly time series generated by the Deterministic and Stochastic DGPs.}
        \label{fig:simulated_group}
\end{figure}

Figure ~\ref{fig:simulated_group} illustrates the examples of simulated daily and hourly time series generated by the two DGPs. Here, we set the $\alpha = \beta = 1$ and $\gamma = 0.2$ for both the DGPs. The lengths of the daily and hourly time series are equivalent to 1096 days and 505 days respectively (one observation more than three seasonal cycles of the highest available seasonality, i.e., 365 * 3 + 1 for the daily data and 168 * 3 + 1 for the hourly data).

Table~\ref{tab:parametersets} summarises the parameter sets ($\alpha$, $\beta$, $\gamma$, $\sigma^2$) used in our experiments. We create 150 time series for each DGP, generating 300 time series of daily and hourly data. As we know the true values of the seasonal, trend, and remainder components of the simulated time series, we calculate the root mean square error (RMSE) for each component by averaging across all datasets and dates. The presence of statistical significance of differences within multiple decomposition methods is assessed using a linear model applied to the squared errors.

Moreover, for the deterministic DGPs we set the \verb|s.window| values of MSTL to $periodic$, whereas for the stochastic DGPs, we use the default \verb|s.window| values of the MSTL, which were obtained through a simulation study (refer to \ref{sec:appendix}).

\begin{table}[!tb]
\caption{The parameter values used by each DGP to simulate time series.}
\centering
\begin{tabular}{lcccc}
\toprule
DGP &$\alpha$ &$\beta$ &$\gamma$ &$\sigma^2$ \\
\midrule
Deterministic  &1 &1 &0.2 &0\\
Deterministic  &1 &1 &0.4 &0\\
Deterministic  &1 &1 &0.6 &0\\
Stochastic  &1 &1 &0.2 &0.025\\
Stochastic  &1 &1 &0.4 &0.050\\
Stochastic  &1 &1 &0.6 &0.075\\
\bottomrule
\end{tabular}
\label{tab:parametersets}
\end{table}

\begin{table}[!ht]
\caption{\label{tab:weeklytable} The RMSE over 150 simulations for each DGP on daily simulated time series data. Bold values indicate results that are significantly different from the MSTL values.}
\centering\small
\begin{tabular}[t]{rrlrrrr}
\toprule{}
$\gamma$ & $\sigma^2$ & Method & Trend RMSE & Weekly RMSE & Yearly RMSE & Remainder RMSE\\
\midrule
\addlinespace[0.3em]
\multicolumn{7}{l}{\textbf{Deterministic DGP}}\\
\hspace{1em}0.2 & 0 & STR & \textbf{0.0174} & \textbf{0.0151} & \textbf{0.0477} & \textbf{0.0514}\\
\hspace{1em}0.2 & 0 & TBATS & \textbf{0.1896} & \textbf{0.0171} & \textbf{0.1822} & \textbf{0.0569}\\
\hspace{1em}0.2 & 0 & PROPHET & \textbf{0.0328} & \textbf{0.0465} & \textbf{0.0423} & \textbf{0.0571}\\
\hspace{1em}0.2 & 0 & MSTL & 0.0623 & 0.0166 & 0.1471 & 0.1429\\
\midrule
\hspace{1em}0.4 & 0 & STR & \textbf{0.0314} & \textbf{0.0291} & \textbf{0.0959} & \textbf{0.1023}\\
\hspace{1em}0.4 & 0 & TBATS & \textbf{0.3752} & \textbf{0.0324} & \textbf{0.3678} & \textbf{0.1081}\\
\hspace{1em}0.4 & 0 & PROPHET & \textbf{0.0524} & \textbf{0.0514} & \textbf{0.0686} & \textbf{0.0805}\\
\hspace{1em}0.4 & 0 & MSTL & 0.0786 & 0.0342 & 0.2471 & 0.2497\\
\midrule
\hspace{1em}0.6 & 0 & STR & \textbf{0.0467} & \textbf{0.0463} & \textbf{0.1438} & \textbf{0.1554}\\
\hspace{1em}0.6 & 0 & TBATS & \textbf{0.5371} & \textbf{0.0526} & \textbf{0.5228} & \textbf{0.1640}\\
\hspace{1em}0.6 & 0 & PROPHET & \textbf{0.0637} & \textbf{0.0631} & \textbf{0.0918} & \textbf{0.1170}\\
\hspace{1em}0.6 & 0 & MSTL & 0.0787 & 0.0556 & 0.3597 & 0.3628\\
\midrule
\addlinespace[0.3em]
\multicolumn{7}{l}{\textbf{Stochastic GDP}}\\
\hspace{1em}0.2 & 0.025 & STR & \textbf{0.0586} & \textbf{0.1492} & \textbf{0.0795} & \textbf{0.1550}\\
\hspace{1em}0.2 & 0.025 & TBATS & \textbf{0.2097} & \textbf{0.1292} & \textbf{0.2004} & \textbf{0.1430}\\
\hspace{1em}0.2 & 0.025 & PROPHET & \textbf{0.0708} & \textbf{0.1542} & \textbf{0.0666} & \textbf{0.1633}\\
\hspace{1em}0.2 & 0.025 & MSTL & 0.1700 & 0.0669 & 0.1638 & 0.1936\\
\midrule
\hspace{1em}0.4 & 0.05 & STR & \textbf{0.0703} & \textbf{0.2623} & \textbf{0.1354} & \textbf{0.2756}\\
\hspace{1em}0.4 & 0.05 & TBATS & \textbf{0.3677} & \textbf{0.2318} & \textbf{0.3506} & \textbf{0.2592}\\
\hspace{1em}0.4 & 0.05 & PROPHET & \textbf{0.0800} & \textbf{0.2629} & \textbf{0.0837} & \textbf{0.2764}\\
\hspace{1em}0.4 & 0.05 & MSTL & 0.1389 & 0.1315 & 0.2439 & 0.2918\\
\midrule
\hspace{1em}0.6 & 0.075 & STR & \textbf{0.0715} & \textbf{0.3811} & \textbf{0.1740} & \textbf{0.3942}\\
\hspace{1em}0.6 & 0.075 & TBATS & \textbf{0.5211} & \textbf{0.3439} & \textbf{0.4934} & \textbf{0.3777}\\
\hspace{1em}0.6 & 0.075 & PROPHET & \textbf{0.0837} & \textbf{0.3810} & \textbf{0.1069} & \textbf{0.3997}\\
\hspace{1em}0.6 & 0.075 & MSTL & 0.1010 & 0.1982 & 0.3482 & 0.4046\\
\bottomrule{}
\end{tabular}
\end{table}

Table~\ref{tab:weeklytable} shows the evaluation summary for the daily simulated time series. With respect to deterministic DGPs, it can be seen that on the Trend, Weekly and Yearly seasonal components, MSTL outperforms the TBATS method in the majority of cases. Whereas, MSTL achieves the best Weekly RMSE values for stochastic DGPs, outperforming all the benchmarks. Also, for stochastic DGPs, MSTL gives better results compared to TBATS on the Trend, Weekly and Yearly seasonal components.

\begin{table*}[!ht]
\caption{\label{tab:hourlytable}The RMSE over 150 simulations for each DGP on hourly simulated time series data. Bold values indicate results that are significantly different from the MSTL values.}
\centering\small
\begin{tabular}[t]{rrlrrrr}
\toprule{}
$\gamma$ & $\sigma^2$ & Method & Trend RMSE & Daily RMSE & Weekly RMSE & Remainder RMSE\\
\midrule
\addlinespace[0.3em]
\multicolumn{7}{l}{\textbf{Deterministic DGP}}\\
\hspace{1em}0.2 & 0 & STR & \textbf{0.0253} & \textbf{0.2736} & \textbf{0.2827} & \textbf{0.0951}\\
\hspace{1em}0.2 & 0 & TBATS & \textbf{0.1004} & \textbf{0.0363} & \textbf{0.0864} & \textbf{0.0737}\\
\hspace{1em}0.2 & 0 & PROPHET & \textbf{0.1170} & \textbf{0.4686} & \textbf{0.6254} & \textbf{0.7872}\\
\hspace{1em}0.2 & 0 & MSTL & 0.0684 & 0.0487 & 0.1400 & 0.1437\\
\midrule
\hspace{1em}0.4 & 0 & STR & \textbf{0.0476} & \textbf{0.2475} & \textbf{0.2697} & \textbf{0.1651}\\
\hspace{1em}0.4 & 0 & TBATS & \textbf{0.1779} & \textbf{0.0734} & \textbf{0.1580} & \textbf{0.1421}\\
\hspace{1em}0.4 & 0 & PROPHET & \textbf{0.1243} & \textbf{0.4377} & \textbf{0.6577} & \textbf{0.7973}\\
\hspace{1em}0.4 & 0 & MSTL & 0.0747 & 0.0885 & 0.2283 & 0.2437\\
\midrule
\hspace{1em}0.6 & 0 & STR & \textbf{0.0613} & \textbf{0.1451} & \textbf{0.1992} & \textbf{0.2318}\\
\hspace{1em}0.6 & 0 & TBATS & \textbf{0.2704} & \textbf{0.1139} & \textbf{0.2436} & \textbf{0.2115}\\
\hspace{1em}0.6 & 0 & PROPHET & \textbf{0.1440} & \textbf{0.4606} & \textbf{0.6520} & \textbf{0.8087}\\
\hspace{1em}0.6 & 0 & MSTL & 0.0858 & 0.1289 & 0.3362 & 0.3614\\
\midrule
\addlinespace[0.3em]
\multicolumn{7}{l}{\textbf{Stochastic GDP}}\\
\hspace{1em}0.2 & 0.025 & STR & \textbf{0.0847} & \textbf{0.0905} & \textbf{0.1253} & \textbf{0.1005}\\
\hspace{1em}0.2 & 0.025 & TBATS & \textbf{0.1324} & \textbf{0.0612} & \textbf{0.1227} & \textbf{0.0890}\\
\hspace{1em}0.2 & 0.025 & PROPHET & \textbf{0.2226} & \textbf{0.4676} & \textbf{0.6483} & \textbf{0.8149}\\
\hspace{1em}0.2 & 0.025 & MSTL & 0.1933 & 0.0952 & 0.1803 & 0.2128\\
\midrule
\hspace{1em}0.4 & 0.05 & STR & \textbf{0.0588} & \textbf{0.8307} & \textbf{0.8351} & \textbf{0.1900}\\
\hspace{1em}0.4 & 0.05 & TBATS & \textbf{0.2215} & \textbf{0.1195} & \textbf{0.2094} & \textbf{0.1716}\\
\hspace{1em}0.4 & 0.05 & PROPHET & \textbf{0.1201} & \textbf{0.4643} & \textbf{0.6344} & \textbf{0.7906}\\
\hspace{1em}0.4 & 0.05 & MSTL & 0.0883 & 0.1779 & 0.2520 & 0.2751\\
\midrule
\hspace{1em}0.6 & 0.075 & STR & \textbf{0.1741} & \textbf{0.2149} & \textbf{0.2529} & \textbf{0.2720}\\
\hspace{1em}0.6 & 0.075 & TBATS & \textbf{0.3081} & \textbf{0.1820} & \textbf{0.2955} & \textbf{0.2573}\\
\hspace{1em}0.6 & 0.075 & PROPHET & \textbf{0.1463} & \textbf{0.4415} & \textbf{0.6168} & \textbf{0.7660}\\
\hspace{1em}0.6 & 0.075 & MSTL & 0.1289 & 0.2577 & 0.3699 & 0.4085\\
\bottomrule{}
\end{tabular}
\end{table*}

Table~\ref{tab:hourlytable} shows the results of all the decomposition techniques for the hourly simulated time series. For the deterministic and stochastic DGPs, we see that the proposed MSTL algorithm gives better RMSE on all components, compared to the PROPHET method. Also, for the \((\gamma, \sigma^2) = \{(0.6, 0.075)\}\) scenario on the Trend component, we that the MSTL achieves the best RMSE.

\subsection{Perturbed real-world data}

We also evaluate the performance of MSTL using real-world time series data. We use the moving block bootstrap (MBB) for real-world time series perturbing, following the procedure introduced in \cite{Bergmeir2016-zk}. Here, we first use MSTL to extract the seasonal, trend and remainder of components of a time series. Assuming these extracted components are the true decomposition of the time series, next we apply the MBB technique to the remainder component of the time series to generate multiple versions of the residual components. Finally, these bootstrapped residual components are added back together with the previously extracted seasonal and trend components to produce new perturbed versions of a time series, where we know the true composition of the series. In our experiments, we use the MBB implementation available through the \verb|MBB| function from the forecast package~\cite{forecastpackagepaper, forecastpackage}.

\begin{figure}[htb]
 \centering
\includegraphics[width=1.00\textwidth]{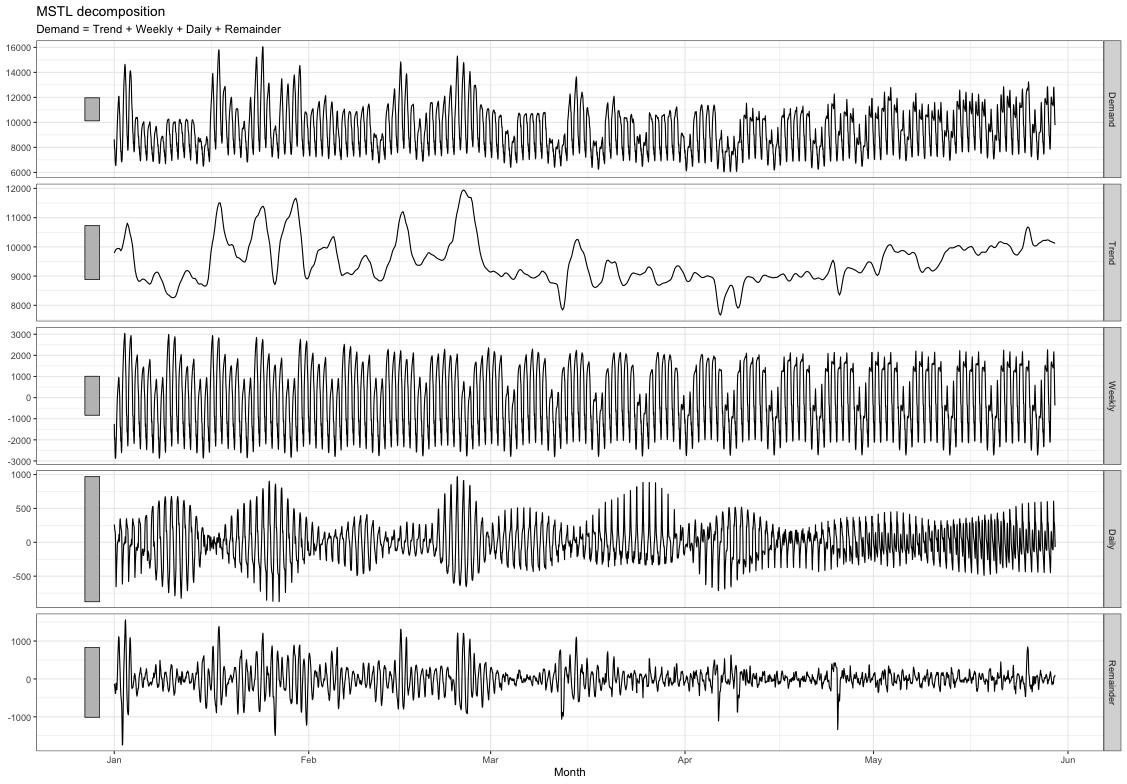}
 \caption{The decomposition of hourly electricity demand in Victoria using MSTL. Each panel represents the original data, the trend, the weekly seasonality, the daily seasonality, and the remainder respectively.}
\label{fig:decomposotion}
\end{figure}

We select the half-hourly electricity consumption in the state of Victoria, Australia, extracted from the \textit{vic\_elec} dataset~\citep{OHara-Wild2020-uq} to generate multiple versions of the time series. As illustrated in Figure~\ref{fig:multiseasonalplot}, this time series has two seasonal patterns, the daily pattern and the weekly seasonal pattern. In our experiments, we first aggregate the half-hourly data to hourly data, and select 149 days starting from 01 January 2012. Figure~\ref{fig:decomposotion} shows the application of MSTL to the hourly electricity demand in Victoria (3601 hourly observations).

\begin{figure}[htb]
 \centering
\includegraphics[width=1.0\textwidth]{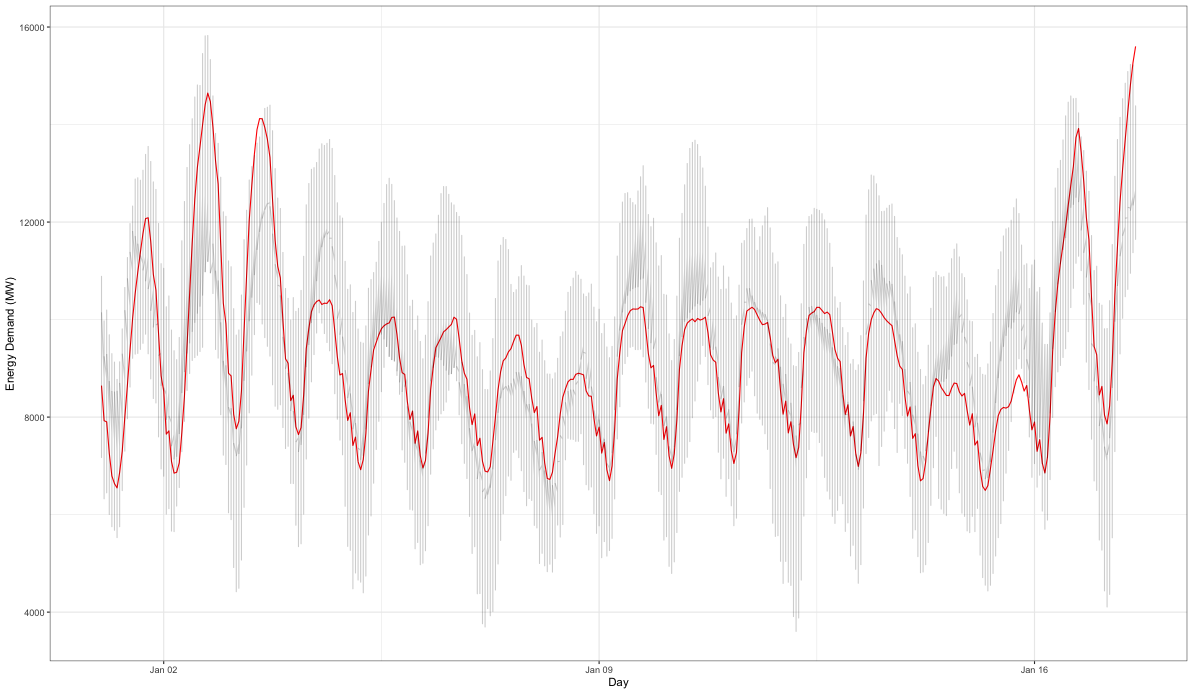}
 \caption{The generation of multiple versions of hourly electricity demand time series, applying the MBB technique to the remainder component of the original time series. The original time series is plotted in red, while the bootstrapped time series are in grey.}
\label{fig:mbb}
\end{figure}

As discussed earlier, we use the MBB technique to generate 100 bootstrapped versions of the hourly electricity demand time series, so we can assess the performance of MSTL on a real-world dataset. Figure~\ref{fig:mbb} illustrates the snippet of original hourly electricity demand time series and the bootstrapped time series.

\begin{table}[!ht]
\caption{\label{tab:realworldresultstable}The RMSE over 100 bootstrapped versions of the hourly electricity demand time series. Bold values indicate results that are significantly different from the MSTL values.}
\centering
\begin{tabular}[t]{crrrr}
\toprule{}
Method & Trend RMSE & Daily RMSE & Weekly RMSE & Remainder RMSE\\
\midrule{}
STR     & \textbf{399.4} & \textbf{408.0} & \textbf{214.8} & \textbf{580.9}\\
TBATS   & \textbf{742.1} & \textbf{348.9} & \textbf{383.1} & \textbf{614.1}\\
PROPHET & \textbf{243.7} & \textbf{371.1} & \textbf{403.9} & \textbf{605.4}\\
MSTL    & 207.6          & 149.2          & 180.5          & 312.7\\
\bottomrule{}
\end{tabular}
\end{table}

Table~\ref{tab:realworldresultstable} summarises the overall performance of MSTL and the benchmarks on the perturbed electricity demand time series. According to Table~\ref{tab:realworldresultstable}, MSTL significantly outperforms STR, TBATS, and PROPHET in estimating all components. Furthermore, Table~\ref{tab:realworldcomputationaltable} provides a summary of the computational cost of MSTL and the benchmarks over the 100 bootstrapped time series. The experiments are run on an Intel(R) i7 processor (1.8 GHz), with 2 threads per core, 8 cores in total, and 16GB of main memory. As shown in Table~\ref{tab:realworldcomputationaltable}, MSTL has the lowest execution time compared to other benchmarks, highlighting the scalability of MSTL to the increasing volumes of time series data. When used for sub-daily time series data, i.e, hourly, half-hourly, the computational efficiency of a decomposition method can be important as those time series are generally longer and contain a higher number of observations.

\begin{table}[!ht]
\caption{\label{tab:realworldcomputationaltable} The total computational cost of the decomposition methods for the electricity demand dataset, measured in seconds.}
\centering
\begin{tabular}{lr}
\toprule
Method & Total time\\
\midrule
STL & 7\\
STR & 612\\
PROPHET & 936\\
TBATS & 2521\\
\bottomrule
\end{tabular}
\end{table}

\section{Conclusions}
Time series datasets with multiple seasonalities are nowadays common in real-world applications. To better understand the variations of such time series, it is often important to decompose the time series into their subcomponents, such as trend, seasonality, and remainder. The existing techniques available for decomposing time series with multiple seasonal cycles are mostly based on complex procedures that can be computationally inefficient for long time series. 

To this end, we have introduced MSTL, a fast time series decomposition algorithm that is capable of handling time series with multiple seasonal cycles. The proposed MSTL algorithm is an extension of the STL decomposition algorithm, which can only extract a single seasonality from a time series. Experimental results on both simulated data and perturbed real-world data have demonstrated that MSTL provides competitive results with lower computational cost in comparison with other state-of-the-art decomposition algorithms, such as STR, TBATS, and PROPHET.

The MSTL algorithm is implemented in the \verb|mstl| function in the \textit{forecast} package~\citep{forecastpackage}, which is available on the Comprehensive R Archive Network (CRAN).

\section{Acknowledgements}
This research was supported by the Australian Research Council under grant DE190100045, and the Australian Centre of Excellence for Mathematical and Statistical Frontiers (Grant CE140100049).

\bibliographystyle{elsarticle-harv}
\bibliography{reference}

\newpage
\appendix
\section{Results of seasonal window simulations}
\label{sec:appendix}
\setcounter{table}{0}
\setcounter{figure}{0}

To determine the default parameter values for the \verb|s.window| parameter values of MSTL, we conduct a series of experiments using a simulation setup similar to Section~\ref{sec:simulation}. Table~\ref{tab:swindowparametersets} summarises the parameter sets ($\alpha$, $\beta$, $\gamma$, $\sigma^2$) used in our experiments. We only use stochastic DGPs to generate time series for this simulation study as real-world time series often contain stochastic time series components.

\begin{figure}
     \centering
     \begin{subfigure}[b]{0.55\textwidth}
         \centering
         \includegraphics[width=\textwidth]{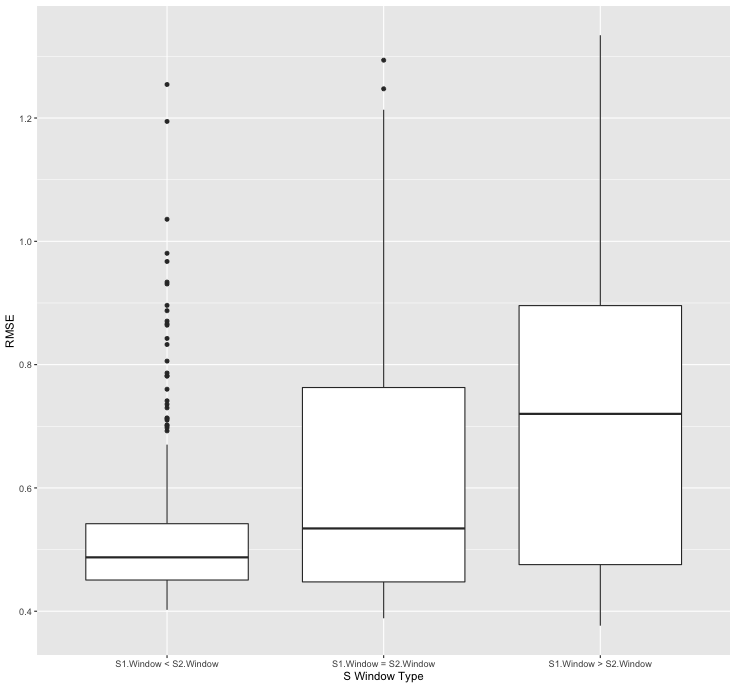}
         \caption{The performance of MSTL across different categories of S1.Window and S2.Window combinations on the weekly dataset.}
         \label{fig:daily_simulated1}
     \end{subfigure}
     \vfill
     \begin{subfigure}[b]{0.55\textwidth}
         \centering
         \includegraphics[width=\textwidth]{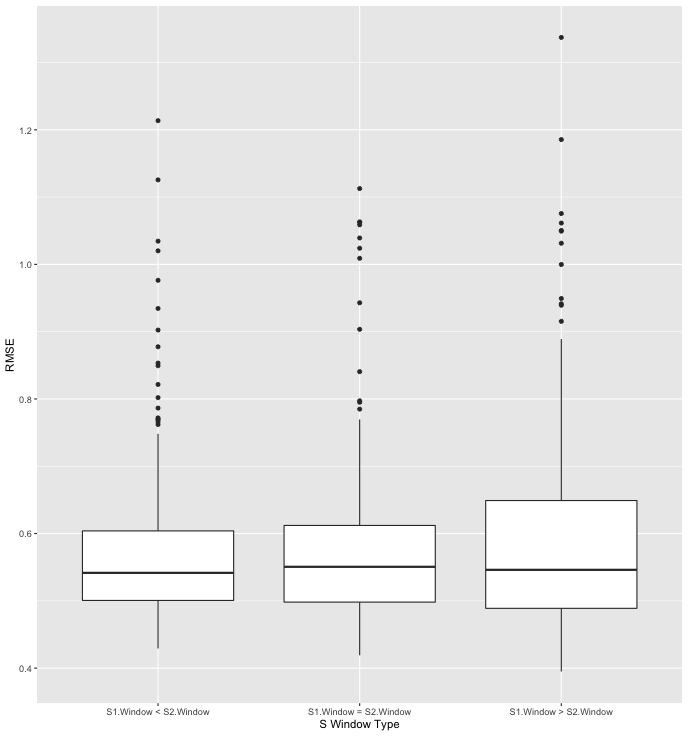}
         \caption{The performance of MSTL across different categories of S1.Window and S2.Window combinations on the hourly dataset.}
         \label{fig:hourly_simulated1}
     \end{subfigure}
        \caption{The RMSE error distribution of MSTL across different categories of S1.Window and S2.Window combinations.}
        \label{fig:boxplot_simulated}
\end{figure}

As we use daily and hourly simulated data, we define the first \verb|s.window| value for the lowest seasonal cycle as S1.Window and the second \verb|s.window| value for the next highest seasonal cycle as S2.Window. For example, S1.Window and S2.window are the \verb|s.window| values for the weekly and yearly seasonal cycles of daily time series and the \verb|s.window| values for the daily and weekly seasonal cycles of hourly time series. Figure~\ref{fig:boxplot_simulated} shows the RMSE error distribution boxplots for the different types of S1.Window and S2.Window combinations on the weekly and hourly datasets. Here, the S1.Window and S2.Window pairs are chosen from the vector $S = (7, 15, 23, 9999)$.

\begin{figure}
     \centering
     \begin{subfigure}[b]{0.90\textwidth}
         \centering
         \includegraphics[width=\textwidth]{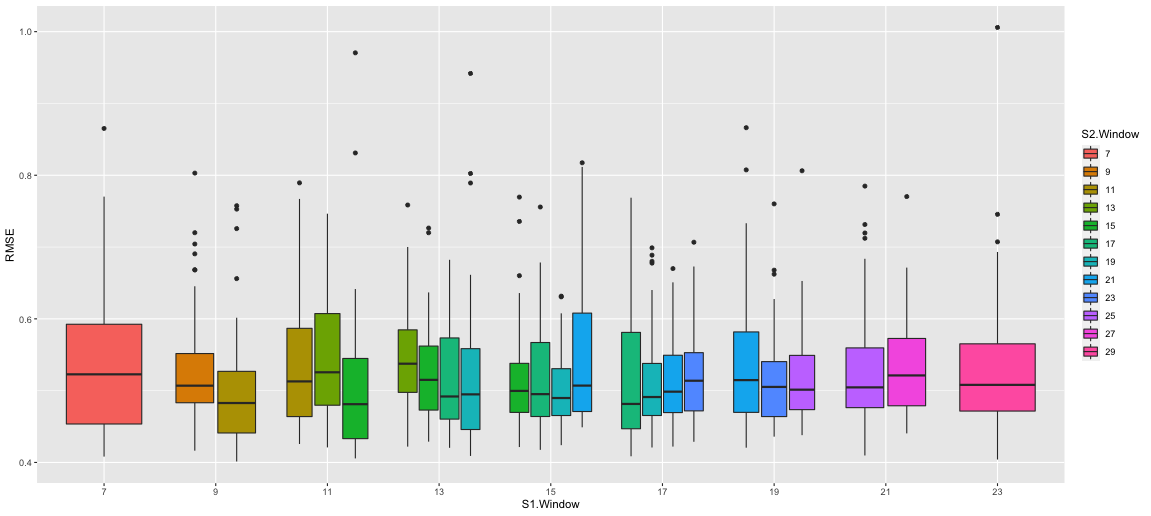}
         \caption{The performance of MSTL across different combinations of S1.Window and S2.Window values on the weekly dataset.}
         \label{fig:daily_simulated2}
     \end{subfigure}
     \vfill
     \begin{subfigure}[b]{0.90\textwidth}
         \centering
         \includegraphics[width=\textwidth]{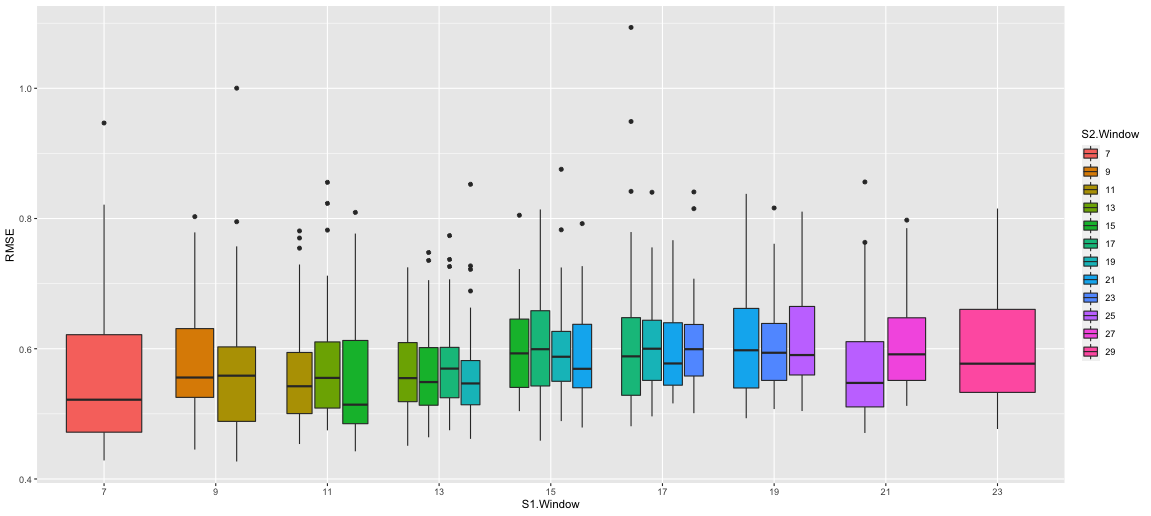}
         \caption{The performance of MSTL across different combinations of S1.Window and S2.Window values on the hourly dataset.}
         \label{fig:hourly_simulated2}
     \end{subfigure}
        \caption{The RMSE error distribution of MSTL across different combinations of S1.Window and S2.Window values.}
        \label{fig:boxplot_all_simulated}
\end{figure}

To identify default parameter values for the \verb|s.window| parameter, we extend this experiments to include more combinations of \verb|s.window| values that satisfy the S1.Window $<$ S2.Window premise. Figure~\ref{fig:boxplot_all_simulated} shows RMSE error distribution boxplots for the different combinations of S1.Window and S2.Window values on the weekly and hourly datasets. The S1.Window and S2.Window pairs are generated from $S = (C+K*i, C + K*i + 1)$, where $C = (7, 9, 11, 13, 15)$ and $K = (0,1,2,3,4,5,6,7)$. Here $i$ represents the seasonal cycle number, i.e., for the first and second seasonality $i =1$ and $i =2$ respectively. Also, we select the smallest odd value from $S$, when choosing a \verb|s.window| value.

\begin{table}[!tb]
\caption{The parameter values used for the seasonal window simulations.}
\centering
\begin{tabular}{lcccc}
\toprule
DGP &$\alpha$ &$\beta$ &$\gamma$ &$\sigma^2$ \\
\midrule
Stochastic  &1 &1 &0.2 &0.025\\
Stochastic  &1 &1 &0.4 &0.050\\
Stochastic  &1 &1 &0.6 &0.075\\
\bottomrule
\end{tabular}
\label{tab:swindowparametersets}
\end{table}

Overall, it can be seen that the S1.Window value of \textbf{11} and the S2.Window value of \textbf{15} gives the best median RMSE for weekly and hourly time series, which is the $C = 7$ and $K = 4$ instance of the above formula. Based on these results, we set the default \verb|s.window| parameter of MSTL using the $S = (C+K*i, C + K*i + 1)$ formula, where $C$ and $K$ values are set to 7 and 4 respectively.

According to Figure~\ref{fig:boxplot_simulated}, the \verb|s.window| combinations that meet the S1.Window $<$ S2.Window condition give the best median RMSE.

\end{document}